\documentclass[twocolumn,showpacs,preprintnumbers,amsmath,amssymb]{revtex4}

\usepackage{epsf}
\usepackage{graphicx}  
\usepackage{dcolumn}   
\usepackage{bm}        

\newcommand{\be}{\begin{equation}}
\newcommand{\en}{\end{equation}}
\newcommand{\bea}{\begin{eqnarray}}
\newcommand{\ena}{\end{eqnarray}}
\newcommand{\cp}{Chaplygin gas}
\begin{document}

\preprint{BNU/065-2005}

\title{Interacting Chaplygin gas }

\author{Hongsheng Zhang\footnote{E-mail address:
hongsheng@bnu.edu.cn} and
  Zong-Hong Zhu\footnote{E-mail address: zhuzh@bnu.edu.cn}}
\affiliation{
 Department of Astronomy, Beijing Normal University, Beijing 100875, China}
\date{ \today}

\begin{abstract}

We investigate a kind of interacting \cp ~model in which the \cp
~plays the role of dark energy and interacts with cold dark matter
particles. We find that there exists a stable scaling solution at
late times with the Universe evolving into a phase of steady state.
Furthermore, the effective equation of state of \cp~ may cross the
so-called phantom divide $w=-1$.  The above results are derived from
continuity equations, which means that they are independent of any
gravity theories. Assuming standard general relativity and a
spatially flat FRW metric, we also find the deceleration parameter
is well consistent with current observations.

\end{abstract}

\pacs{ 98.80. Cq } \keywords{\cp, dark energy}

\maketitle

\section{Introduction}

 The existence of dark energy is one of the most significant cosmological discoveries
 over the last century \cite{acce}. However, although fundamental for
our understanding of the Universe, its nature remains a completely
open question nowadays. Various models of dark energy have been
proposed, such as a small positive
 cosmological constant, quintessence, k-essence, phantom, holographic dark energy,
 etc., see \cite{review} for
 recent reviews with fairly complete lists of references of different dark energy
models.

 Recently the so-called \cp, also dubbed quartessence, was suggested as a
candidate of a unified model of dark energy
 and dark matter \cite{cp}.  The \cp~ is characterized by an exotic equation
of state
 \be
 p_{ch}=-A/\rho_{ch},
 \label{state}
 \en
 where $A$ is a positive constant. The above equation of state leads to
 a density evolution in the form
 \be
 \rho_{ch}=\sqrt{A+\frac{B}{a^6}},
 \label{evo}
 \en
 where $B$ is an integration constant.
The attractive feature of the model is that it naturally unifies
both dark energy and dark matter. The reason is that, from
(\ref{evo}), the \cp~ behaves as dust-like matter at early stage and
as a cosmological constant at later stage. Some possible motivations
for this model from the field theory points of view are
 investigated in ~\cite{fields}. The Chaplygin gas emerges as an effective
fluid associated with $d$-branes~\cite{brane} and can also be
 obtained from the Born-Infeld action~\cite{bi}. Recently, the original \cp model was
generalized, with possible observational constraints on these
generalized models presented in Ref. \cite{many}. For example in the
generalized Chaplygin gas (GCG) approach \cite{gcg}, the equation of
state to describe the background fluid is generalized to
 \be
 p_{ch} = - {A \over \rho_{ch}^{\alpha}},
 \label{gcgstate}
 \en
 and the corresponding evolution of the scale factor is given by
 \be
 \rho_{ch}=  \left[A + {B \over a^{3 (1 + \alpha)}}\right]^{1 \over 1
 + \alpha}.
 \en
 From the above equations, it is clear that when $\alpha=1$ the GCG model recovers the original \cp ~model.
  This approach has been thoroughly investigated
 for its impact on the 0th order
cosmology, i.e., the cosmic expansion history (quantified by the
Hubble parameter $H[z])$ and corresponding spacetime-geometric
observables. An interesting range of models was found to be
consistent with SN Ia data \cite{sn}, CMB peak locations \cite{cmb}
and dimensionless coordinate distances to type Ia supernovae
\cite{zhzh}. There seems to be, however, a flaw in unified dark
matter (UDM) models that manifests
 itself only on small (Galactic) scales and that has not been revealed by the
 studies involving only background tests. In Ref. \cite{antiudm},
  it is found that GCG model produces oscillations or exponential blowup of the matter
   power spectrum inconsistent with observations. In fact, from this analysis, 99.999 \% of
   previously allowed parameter of GCG model has been excluded (see, however, \cite{bbc}).

            Hence we may turn to a model with \cp~and dark matter.
        Although non-minimal coupling between the dark energy and ordinary
matter fluids  is strongly restricted by the  experimental tests in
the solar system \cite{will}, due to the unknown nature of the dark
matter as part of the background, it is possible to have
non-gravitational interactions between the dark energy and the dark
matter components, without conflict with the experimental data. In
this paper we investigate some physical properties of an interacting
\cp~ model. Here, by considering an interaction term between the
\cp~fluid and dark matter particles similar to those studied in the
context of quintessence scenarios \cite{quph}, we investigate
dynamical aspects of this interacting \cp~ model.

  Following the more accurate data a more dramatic result
  appears:
  the recent analysis of the type Ia supernovae data
  indicate that the time varying dark energy gives a better
  fit  than a cosmological constant, and in particular, the equation of state parameter
   $w$ (defined as the ratio of
 pressure to energy density) crosses $-1$ at some low redshift region from above to below
 \cite{vari}, where $w=-1$ is the equation of state for the
 cosmological constant. It deserves to note that there are other independent
  fittings imply the probability that current $w<-1$ except for
  supernovae data \cite{zhzh2}.  The dark energy with $w<-1$ is called phantom dark
 energy~\cite{call}, for which all energy conditions are
 violated.
  To obtain $w <-1$, scalar field with a negative kinetic term,
  may be a simplest realization \cite{phantom}. However,
 the equation of state of phantom scalar field is  always
 less than $-1$  and can not cross $-1$. Also it has been shown
 that the equation of state cannot cross $-1$ in the k-essence
 model of dark energy under some reasonable
 assumptions~\cite{Vik}. Some dark energy models which contain a negative-kinetic scalar field and a normal
scalar field  have been considered in \cite{guozk2}; in these models
crossing the border $w=-1$ can be realized. Some different
suggestions on this crossing behavior are presented in \cite{And}.

 It has been pointed out that \cp~ model can be described by
 a quintessence filed with well-connected potential \cite{cp}.
 So a model with mutually independent \cp~ and dark matter
 is essentially a special quintessence model. The \cp~, here as dark
 energy, can not cross the phantom divide like quintessence. In this
 paper we shall see that an interaction term can realize this
 crossing naturally. At the same time we obtain a scaling solution:
 It may also shed light on the coincidence problem. Another
 interesting result is that the scaling solution inevitably leads to
  the steady state Universe \cite{bondi},
   which had been suggested many years ago
  but soon surpassed by expanding Universe, as the final state of our
  model.

     We present our model in details in the next section and
     some observational
predictions of this scenario and a comparison with recent
observational data are also briefly discussed.
     Our conclusions and discussions appear in the last section.

\section{the model }
 We consider the original \cp, whose pressure and energy density satisfy the
 relation, $p_{ch}=-A/\rho_{ch}$. By assuming the cosmological principle
 the continuity equations are written as
 \be
 \dot{\rho}_{ch}+3H \gamma_{ch} \rho_{ch}=-\Gamma,
 \label{1st conti}
 \en
 and
 \be
 \dot{\rho}_{dm}+3H \gamma_{dm} \rho_{dm}=\Gamma,
 \label{2nd conti}
 \en
 where the subscript $dm$ denotes dark matter, $H$ is the Hubble
 parameter and $\gamma$ is defined as
 \be
 \gamma=1+\frac{p}{\rho}=1+w,
 \en
 in which $w$ is the parameter of the state of equation,
  and $\gamma_{dm}=1$ throughout the evolution of the Universe,
 whereas $\gamma_{ch}$ is a variable.

 $\Gamma$ is the
 interaction term between \cp~and dark matter. Since there does not
exist any microphysical hint on the possible nature of a coupling
between dark matter and \cp~(as dark energy), the interaction terms
between dark energy and dark matter are rather arbitrary in
literatures \cite{inter}. Here
 we try to present a possible origin from fundamental field theory for
 $\Gamma$ (see \cite{maia} for a thermodynamic
 discussion on $\Gamma$).

 Whereas we are still lacking a
complete formulation of unified theory of all interactions
(including gravity, electroweak and strong), there at present is at
least one very hopeful candidate, string/M theory. Although the
recent developments in string theory, assisted by the discovery of
the power of duality, have greatly improved our understanding of it,
the theory is still not known in a way that would enable us to ask
the questions about space-time in a general manner, say nothing of
the properties of realistic particles. Instead, we have to either
resort to the effective action approach which takes into account
stringy phenomena in perturbation theory, or we could study some
special classes of string solutions which can be formulated in the
non-perturbative regime. But the latter approach is available only
for some special solutions, most notably the BPS states or nearly
BPS states in the string spectrum: They seems to have no relation to
our realistic Universe. Especially, there still does not exist a
non-perturbative formulation of generic cosmological solutions in
string theory. Hence nearly all the investigations of realistic
string cosmologies have been carried out essentially in the
effective action range \cite{stringcos}. Note that the departure of
string-theoretic solutions away from general relativity is induced
by the presence of additional degrees of freedom which emerge in the
massless string spectrum. These fields, including the scalar dilaton
field, the torsion tensor field, and others, couple to each other
and to gravity non-minimally, and can influence the dynamics
significantly. Thus such an effective low energy string theory
deserve research to solve the dark energy problem. There a special
class of scalar-tensor theories of gravity is considered to avoid
singularities in cosmologies in \cite{st}. The action is written
below,
 \bea
 S_{st}=\int d^4x \sqrt{-g} \left[\frac{1}{16\pi G}R-
 \frac{1}{2}\partial_{\mu} \phi \partial^{\mu} \phi +\right.
 \nonumber \\ \left.
 \frac{1}{q(\phi)^2}
 L_m(\xi, \partial \xi, q^{-1}g_{\mu\nu})\right],
 \label{staction}
 \ena
  where $G$ is the Newton gravitational constant, $\phi$ is a scalar
  field, $L_m$ denotes Lagrangian of matter , $\xi$ represents
  different matter degrees of matter fields, $q$ guarantees the coupling strength
  between the matter fields and the dilaton. With action
  (\ref{staction}), the interaction term can be written as follow \cite{st},
  \be
  \Gamma=H\rho_m \frac{d\ln q'}{d\ln a}.
  \en
  Here we introduce new variable $q(a)'\triangleq
  q(a)^{(3w_n-1)/2}$, where $a$ is the scale factor in standard
  FRW metric.   By assuming
  \be
  q'(a)=q_0e^{3\int c(\rho_m+\rho_{\xi})/\rho_m d\ln a  },
  \en
  where $\rho_m$ and $\rho_{\xi}$ are the densities of
  matter and the scalar field respectively, one arrive at the interaction term,
  \be
  \Gamma=3Hc(\rho_m+\rho_\xi),
  \en
  which is just the coupling form studied in contexts of quintessence and phantom
  dark energy models \cite{quph}. Moreover the \cp~ can be view as a
  scalar field with proper potential in cosmological models \cite{cp}.
  So it may be reasonable to phenomenologically introduce such an interaction term between
  \cp ~and dark matter.

  Now return to the equation set (\ref{1st conti}) and
  (\ref{2nd conti}). Set
  $x=-\ln(1+z)$, $\Gamma=3Hc(\rho_{ch}+\rho_{dm})$,
  $u=(3H_0^2)^{-1}\kappa^2 \rho_{dm}$, $v=(3H_0^2)^{-1}\kappa^2
  \rho_{ch}$, $A'=A(3H_0^2)^{-2}\kappa^4$, where $H_0$ takes the value of
  present Hubble parameter, $\kappa$ is the Newton gravitational
  constant and $c$ is a constant without dimension.
   Eqs. (\ref{1st conti}) and
  (\ref{2nd conti}) reduce to
  \be
  \frac{du}{dx}=-3u+3c(u+v),
  \label{auto1}
  \en
  \be
  \frac{dv}{dx}=-3(v-A'/v)-3c(u+v).
  \label{auto2}
  \en
  We note that the variable time does not appear in the dynamical system
  (\ref{auto1}) and (\ref{auto2}) because time has been completely
  replaced by redshift $x=-ln(1+z)$. The critical points of
  dynamical system (\ref{auto1}) and (\ref{auto2}) are given by
  \be
  \frac{du}{dx}=\frac{dv}{dx}=0.
  \en
  The solution of the above equation is
  \bea
  u_c=\frac{c}{1-c}v_c,
 \label{uc}
  \\
  v_c^2=(1-c)A'.
  \label{vc}
  \ena
  We see the final state of the model contains both \cp ~and dark matter
  of constant densities if the singularity is stationary. The final
  state contents perfect cosmological principle: the Universe is
   homogeneous and isotropic in space, as well as constant in time.
   Physically $\Gamma$ in (\ref{2nd conti}) plays the role of matter
   creation term $C$ in the theory of steady state universe at the
   future time-like infinity.
  Recall that $c$ is the coupling constant, may be positive or negative,
  corresponds the energy to transfer from \cp~ to dark matter or
  reversely. $A'$ must be a positive constant, which denotes the
  final energy density if $c$ is fixed. Also we can derive an
  interesting and simple relation between the static energy density
  ratio
  \be
  c=\frac{r_s}{1+r_s},
  \en
  where
  \be
  r_s=\lim_{z\to -1}\frac{\rho_{dm}}{\rho_{ch}}.
  \en
  To investigate the properties of the dynamical system in the
  neighbourhood of the singularities, impose a perturbation
  to the critical points,
  \bea
  \frac{d(\delta u)}{dx}=-3\delta u+3c(\delta u+\delta v),
  \label{linear1}
  \\
  \frac{d(\delta v)}{dx}=-3(\delta v+\frac{A'}{v_c^2}\delta v)-3c(\delta
  u+\delta v).
  \label{linear2}
  \ena
  The eigen equation of the above linear dynamical system $(\delta u,
  \delta v)$ reads
  \be
  (\lambda/3)^2+(2+\frac{1}{1-c})\lambda/3+2-2c^2=0,
  \label{eigen}
  \en
  whose discriminant is
  \be
  \Delta=[(1-c)^4+(3/2-c)^2]/(1-c)^2\geq 0.
  \en
  Therefore both of the two roots of eigen equation (\ref{eigen}) are real, consequently
    centre and focus singularities can not appear.
  Furthermore only $r_s\in (0,\infty)$, such that $c\in (0,1)$, makes
  physical sense. Under this condition it is easy to show that both
  the two roots of (\ref{eigen}) are negative. Hence the two
  singularities are stationary. However it is only the property
  of the linearized system (\ref{auto1}) and (\ref{auto2}),
  or  the property of orbits of the neighbourhoods
  of the singularities, while global Poincare-Hopf theorem requires that the
  total index of the singularities equals the Euler number of the
  phase space for the non-linear system (\ref{linear1}) and (\ref{linear2}). So there exists other
  singularity except for the two nodes. In fact it is a non-stationary saddle point at
  $u=0,~v=0$ with index $-1$. This singularity has been omitted in solving
  equations (\ref{auto1}) and (\ref{auto2}). The total index of the three singularities
  is $1$, which equals the Euler number of the phase space of this plane dynamical
  system. Hence there is no other singularities in this system.
    From these discussions we conclude that the global outline of
  the orbits of this non-linear dynamical system (\ref{auto1}) and (\ref{auto2})
   is similar to the electric fluxlines of two
  negative point charges.
    Here we plot figures 1 and 2 to show the properties of evolution of
    the Universe controlled by the dynamical system (\ref{auto1}) and
    (\ref{auto2}). As an example we set $c=0.2,~~A'=0.9$ in all the figures except figure \ref{dece}.

  \begin{figure*}
\centering
\includegraphics[totalheight=2.2in, angle=0]{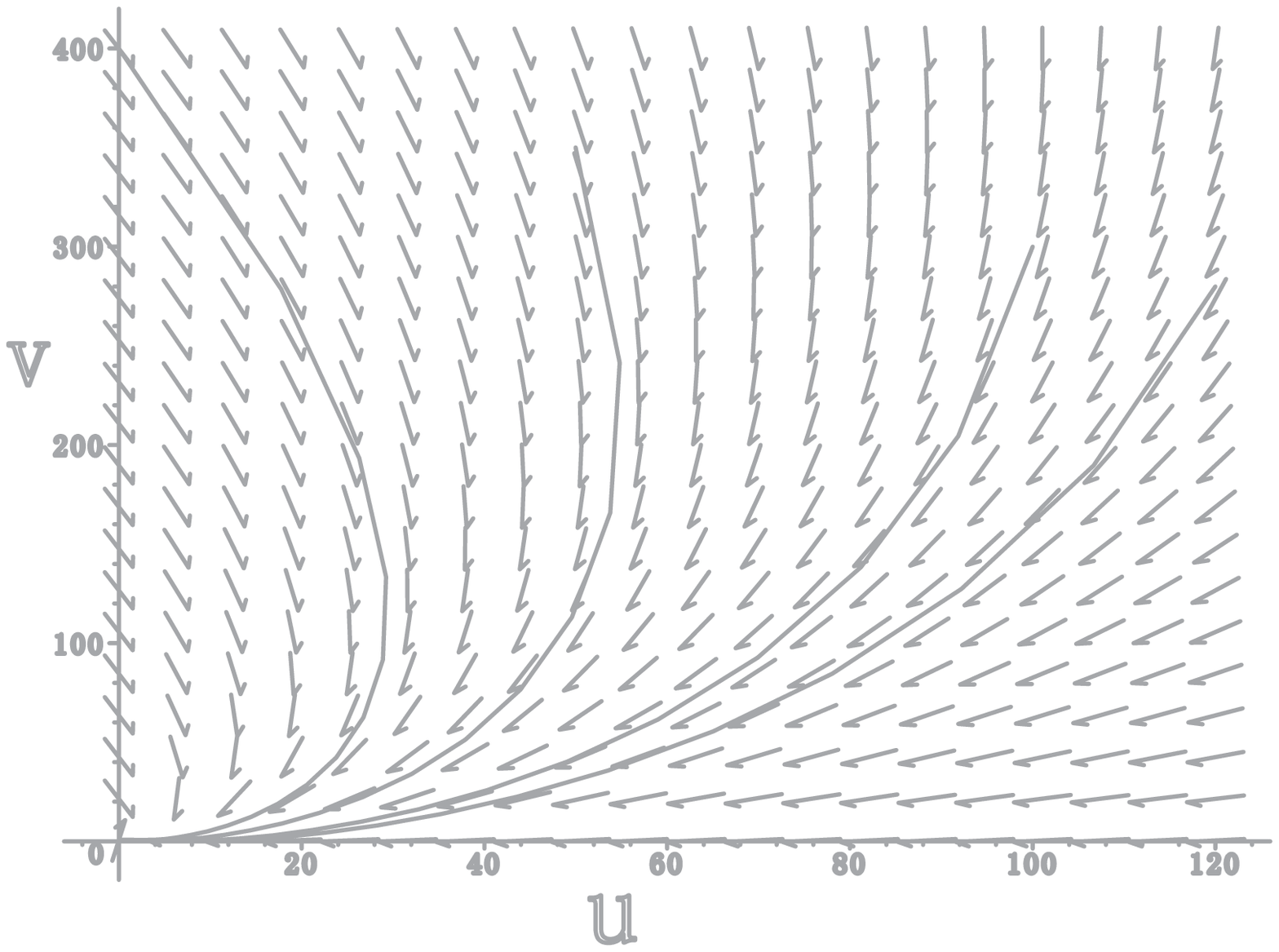}
\includegraphics[totalheight=2.2in, angle=0]{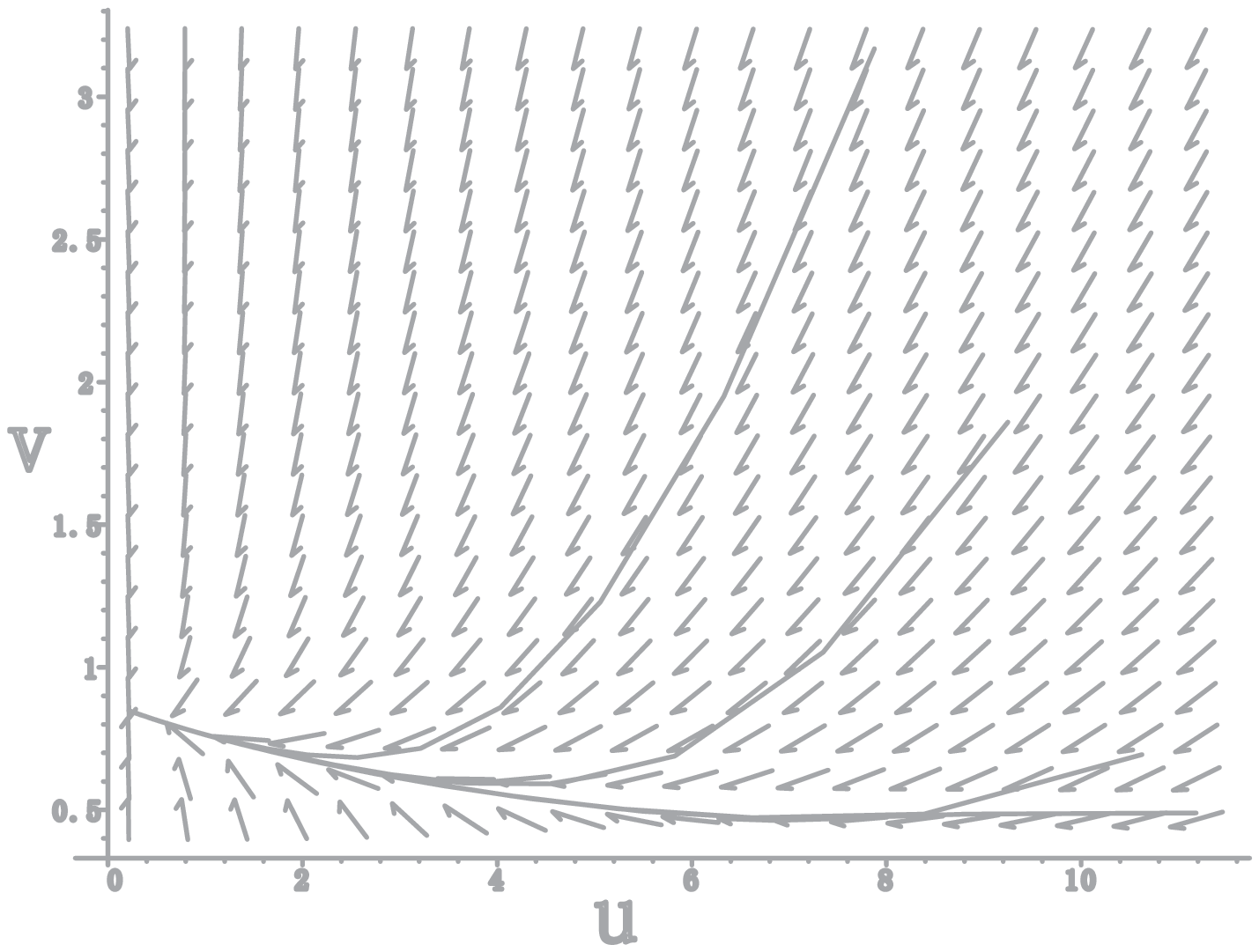}
\caption{The plane v versus u. {\bf{(a)}} We consider the evolution
of the universe from redshift $z=e^2-1$. The initial condition is
taken as $u=0,~v=400$; $u=50,~v=350$; $u=100,~v=300$; $u=120,~v=280$
on the four orbits, from the left to the right, respectively. It is
clear that there is a stationary node, which attracts most orbits in
the first quadrant. At the same time the orbits around the
neibourhood of the singularity is not shown clearly. {\bf{(b)}}
Orbit distributions around the node $u_c=v_c
c/(1-c),~v_c=\sqrt{(1-c)A'}$.}
 \label{globalgreen}
 \end{figure*}

 \begin{figure}
\centering
\includegraphics[totalheight=3in, angle=-90]{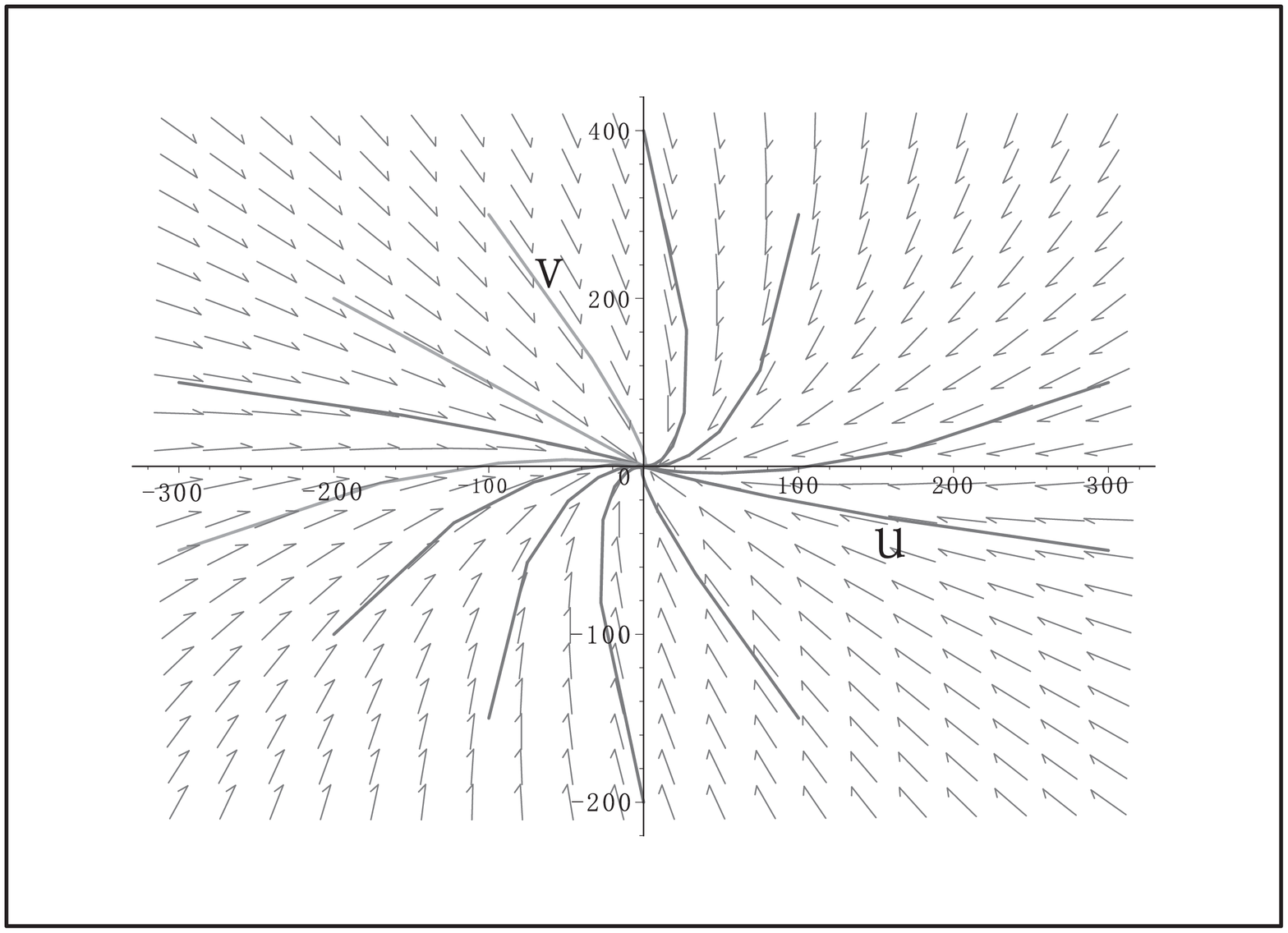}
\includegraphics[totalheight=3in, angle=-90]{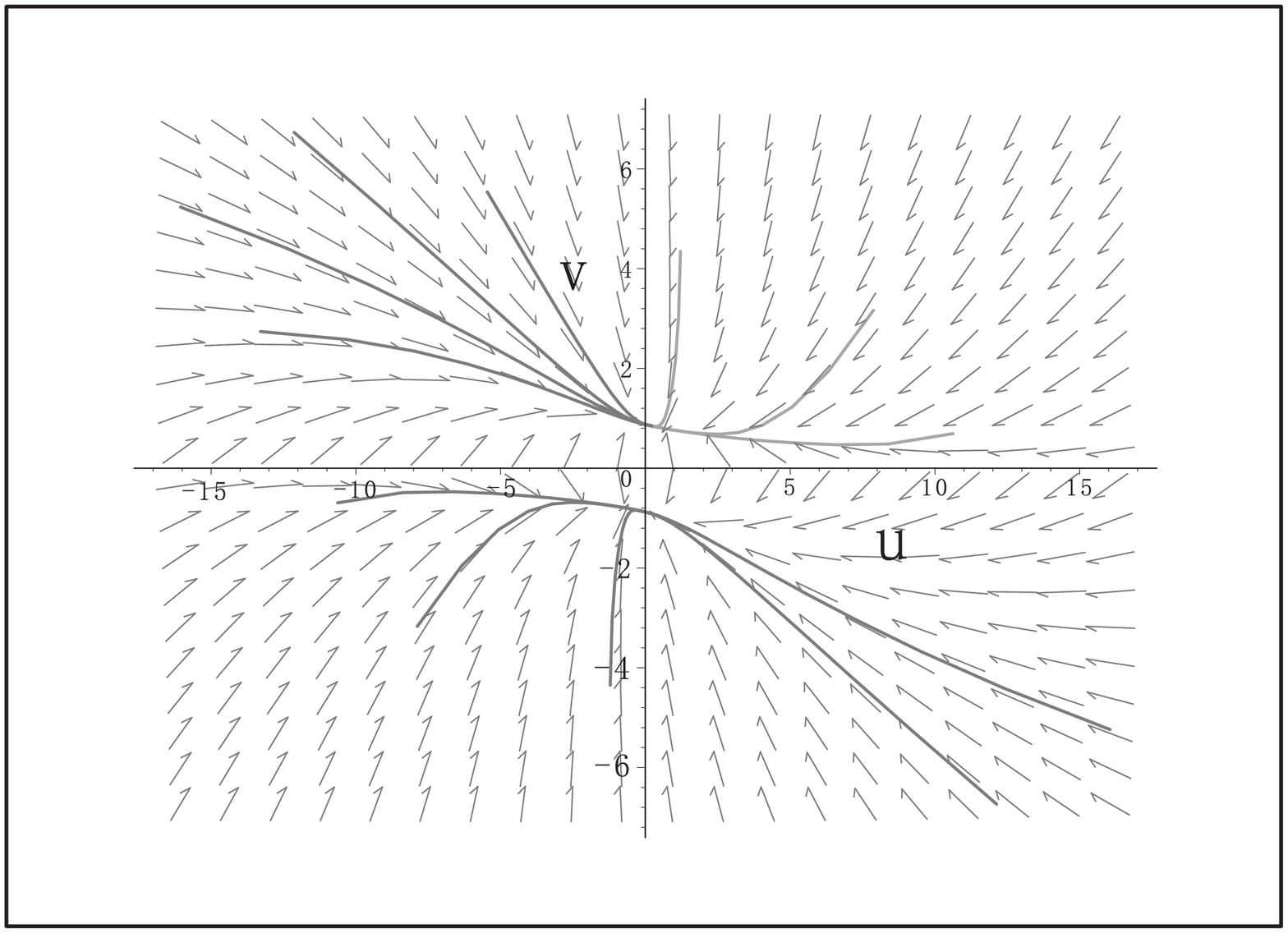}
\caption{The plane v versus u. {\bf{(a)}}    To show the global
properties of dynamical system (\ref{auto1}) and (\ref{auto2}) we
have to include some ``unphysical '' initial conditions, such as
$u=-100,~v=-300$, except for physical initial conditions which have
been shown in figure \ref{globalgreen}. {\bf{(b)}} Orbits
distributions around the nodes. The two nodes
 $u_c=v_c c/(1-c),~v_c=\sqrt{(1-c)A'}$ and $u_c=v_c c/(1-c),~v_c=-\sqrt{(1-c)A'}$
 keep reflection symmetry about the original point. Just as we have analyzed, we see
 that the orbits of this dynamical system are similar to the electric fluxlines of two
  negative point charges.}
 \label{global}
 \end{figure}

\begin{figure}
 \centering
 \includegraphics[totalheight=1.95in, angle=0]{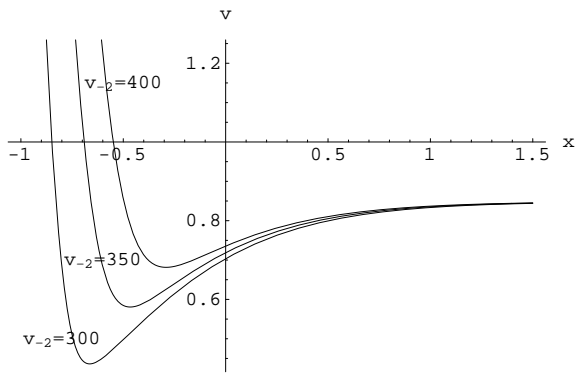}
 \caption{v versus x. The initial condition of the three curves is
 $u(-2)=0,~v(-2)=400$;
  $u(-2)=50,~v(-2)=350$; $u(-2)=100,~v(-2)=300$, respectively. Obviously the energy density
  of \cp~  rolls down and then climbs up in some low redshift region. So the \cp
  ~dark energy can cross the phantom divide $w=-1$ in a fitting where the
  dark energy is treated as an independent component to dark matter.   }
 \label{figv}
 \end{figure}

\begin{figure}
 \centering
 \includegraphics[totalheight=2in, angle=0]{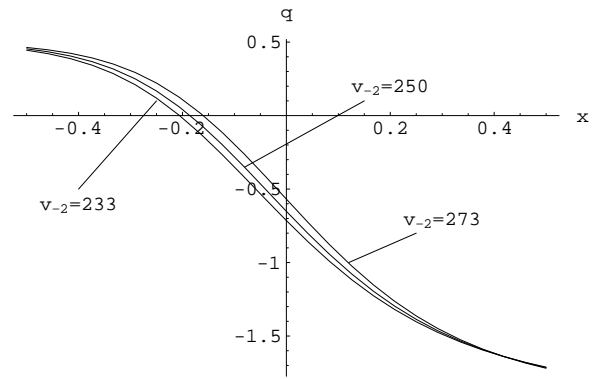}
 \caption{q versus x. The initial condition of the three curves is
 $u(-2)=0,~v(-2)=273$;
  $u(-2)=15,~v(-2)=250$; $u(-2)=25,~v(-2)=233$, respectively. Evidently the
  deceleration parameter $q$
  of \cp~  rolls down and crosses $q=0$ in some low redshift region. The transition
  from deceleration phase to acceleration phase occurs at $z=0.18;~z=0.21;~z=0.23$
  to the curves $u(-2)=0,~v(-2)=273$;
  $u(-2)=15,~v(-2)=250$; $u(-2)=25,~v(-2)=233$, respectively.  One
  finds $-q\thickapprox 0.5\sim 0.6$ at $z=0$, which is well consistent with  observations.   }
 \label{dece}
 \end{figure}

  Further, to compare with observation data we need the explicit
  forms of $u(x)$ and $v(x)$, especially $v(x)$, since we have set
  $\gamma_{dm}=1$ but $\gamma_{ch}$ is not a constant. We need
  the properties of $\gamma_{ch}$ in our model, which is contained in
  $v(x)$, to compare with observations. Eliminate $u(x)$ by using
  (\ref{auto1}) and (\ref{auto2}) we derive
  \bea
  \frac{1}{3c}\frac{d^2v}{dx^2}+[1+(1+A'/v^2)/c]\frac{dv}{dx}+3cv+ \nonumber \\
  3(1-c)\left\{v+\left[\frac{dv}{dx}+3(v-A'/v)\right]/(3c)\right\}=0,
  \ena
  which has no analytic solution. We show some numerical solutions
  in figure \ref{figv}. We find that for proper region of parameter
  spaces, the effective equation of state of \cp~ crosses the
  phantom divide successfully. We have no analytical result on this
  crossing phenomena yet in the present stage.

  Up to now all of our results do not depend on Einstein field
  equation. They only depend on the most sound principle in physics, that is, the
  continuity principle, or the energy conservation law.  Different gravity theories,
   such as standard general relativity, brane-induced gravity, $1/R$ gravity,
   and Lovelock gravity \cite{severalg} or
   cosmology of modified Friedmann equation such as Cardassian
   cosmology \cite{card}, correspond to different constraints imposed on our
   previous discussions.
    Our improvements show how far we can reach without information of dynamical
    evolution of the Universe.

    The most significant parameters from the viewpoint of
  observations is the deceleration parameter $q$, which carries the total
  effects of cosmic fluids. From now on we introduce the Friedmann
  equation of the standard general relativity.
  As a simple case we study the evolution of $q$ in
  a spatially flat frame. So $q$ reads

  \be
  q=-\frac{\ddot{a}a}{\dot{a}^2}=\frac{1}{2}\left(\frac{u+v-3A'/v^2}
  {u+v}\right),
  \en
  and density of \cp~ $u$ and density of dark matter $v$ should
  satisfy
  \be
  u(0)+v(0)=1.
  \label{con}
  \en
  And then Friedmann equation ensures the spatial flatness in the
  whole history of the Universe.
  Before analyzing the evolution of $q$ with redshift,
   we first study its asymptotic behaviors. When $z\to \infty$, $q$
 must go to $1/2$ because both \cp~ and dark matter behave like dust
 , while when $z\to -1$ $q$ is determined by
 \be
  \lim_{z\to -1} q=\frac{1}{2}\left(\frac{u_c+v_c-3A'/v_c^2}
  {u_c+v_c}\right).
  \en
 One can finds the parameters $c=0.2,~A'=0.9$ are difficult to
  content the previous constraint Friedmann constraint ($\ref{con}$). Here we carefully
  choose a new set of parameter which satisfies Friedmann constraint
  (\ref{con}), say, $A'=0.4,~c=0.06$. Therefore we obtain
 \be
 \lim_{z\to -1} q=-1.95,
 \en
  by using (\ref{uc}) and (\ref{vc}).
  Then we plot figure \ref{dece} to clearly display the evolution of
  $q$. One can check $u(0)=0.25,~v(0)=0.75;~u(0)=0.28,~v(0)=0.72;
  ~u(0)=0.3,~v(0)=0.7$, respectively on the curves
  $v(-2)=273;~v(-2)=250;~v(-2)=233$. One may find an interesting
  property of the deceleration parameter displayed in figure
  \ref{dece}: The bigger the proportion of the dark energy, the
  smaller the absolute value of the deceleration parameter. The
  reason roots in the extraordinary state of \cp~(\ref{state}), in
  which the pressure $p_{ch}$ is inversely proportional to the energy density $\rho_{ch}$.

 Also we note that maybe an FRW Universe with non-zero
  spatial curvature fits  deceleration parameter
  better than spatially flat FRW Universe. This point deservers to
  research further.


\section{Conclusion and discussion}

 We present a phase-space analysis of the evolution for a
  FRW universe driven by an interacting mixture of dark matter and
 \cp. In the absence of
interaction, there exist no scaling solutions because the state of
equation of \cp~ decreases with scale factor while state of equation
of dark matter keeps a constant. Hence we study the existence and
stability of the cosmological scaling solutions with interaction
between \cp ~and dark matter. For the interaction term
$\Gamma=3\,c\,H(\rho_{ch}+\rho_{dm})$, inspired by low energy
effective string theory and scalar-tensor gravity theory, we find
stationary scaling solutions for reasonable initial conditions.
Further more this approach leads to an impressive result: The
equation of state of \cp~traverses the phantom divide $w=-1$, which
is favored by recent fittings with current type Ia supernovae data.
At the same time,
 different from phantom model, there is no singularity in the future
 in our model. On the contrary the final state of our model is
 steady state Universe, in which \cp~ ensures the continuous
 production of matter (dust) through the interaction term.  The analytical
 researches on this crossing behavior deserve to investigate in the future work.
      We also
 calculate the deceleration parameter of this model in frame of
 spatially flat FRW cosmology. The result is consistent with the observation
 data.

     At last, as a phenomenological model, we must constrain it by
     observational data.  We see that this model is so simple that it is
     fully parameterized by two parameters: $c$ determines the
  final ratio of \cp ~and dark matter, and $A$ governs the final total
  energy density of the Universe. We may consider the observational
  constraints on the parameter space arising from
  observation data from different observations, such as supernovae,
  CMB, X-rays, gravitational lensing effects or combination of these
  data in future work.

 {\bf Acknowledgments.}
 We thank Z. Guo and J. S. Alcaniz  for helpful discussions. This work was supported by
  the National Natural Science Foundation of China
    , under Grant No. 10533010, and by SRF for ROCS, SEM of China.


\begin{thebibliography}{99}

\bibitem{acce}
  A. G. Riess et al.,
  Astron. J. 116, 1009 (1998), astro-ph/9805201;
  S. Perlmutter et al.,
  Astrophys. J. 517, 565 (1999), astro-ph/9812133.
\bibitem{review}
 T. Padmanabhan, Phys. Rept. 380, 235 (2003), hep-th/0212290;
 P. J. E.Peebles and B. Ratra,
 Rev. Mod. Phys. 75, 559 (2003), astro-ph/0207347;
 V. Sahni, astro-ph/0403324.
\bibitem{cp}
  A. Kamenshchik, U. Moschella and V. Pasquier, Phys. Lett.
  {\bf B511} (2001) 265;
\bibitem{fields}
  N. Bilic, G.B. Tupper and  R.D. Viollier, Phys. Lett.
  {\bf B535} (2002) 17;
  N. Bilic, G.B. Tupper and R.D. Viollier, astro-ph/0207423.
\bibitem{brane}
 M. Bordemann and J. Hoppe,  Phys. Lett. {\bf B317} (1993) 315;
 J.C. Fabris, S.V.B. Gonsalves and P.E. de Souza,
 Gen. Rel. Grav. {\bf 34} (2002) 53.
\bibitem{bi}
  M.C. Bento, O. Bertolami and A.A. Sen, Phys. Lett.
  {\bf B575} (2003) 172.


\bibitem{many}
 Z. Guo and Y. Zhang, astro-ph/0509790;
 Z. Guo and Y. Zhang, astro-ph/0506091;
  G. Sethi, S. Singh, P. Kumar, D. Jain and A. Dev, astro-ph/0508491;
  A. Sen and R. Scherrer, astro-ph/0507717;
  O. Bertolami and P. Silva, astro-ph/0507192;
  M. Makler, B. Mota and M. Reboucas, astro-ph/0507116;
  B. Mota, M. Makler and M. Reboucas, astro-ph/0506499;
  M. Mak and T. Harko, Phys.Rev. D71 (2005) 104022, gr-qc/0505034;
  O. Bertolami, astro-ph/0504275;
  W. Zimdahl, J. Fabris, gr-qc/0504088;
  N. Bilic, G. Tupper and R. Viollier, astro-ph/0503428;
  R. Colistete Jr. and J. Fabris, Class.Quant.Grav. 22 (2005) 2813,
  astro-ph/0501519;
  D. Liu and X. Li, Chin.Phys.Lett. 22 (2005) 1600,
  astro-ph/0501115;
  L. Chimento, R. Lazkoz, Phys.Lett. B615 (2005) 146,
  astro-ph/0411068;
  U. Debnath, A. Banerjee and S. Chakraborty, Class.Quant.Grav.
  21 (2004) 5609, gr-qc/0411015;
  F. Perrotta, S. Matarrese and M. Torki, Phys.Rev. D70 (2004)
  121304;
   V.~Gorini, A.~Kamenshchik, U.~Moschella, V.~Pasquier and A.~Starobinsky,
  Phys.\ Rev.\ D {\bf 72}, 103518 (2005)
  astro-ph/0504576;
  R. Colistete Jr., J. C. Fabris and S.V.B. Goncalves,
  Int.J.Mod.Phys. D14 (2005) 775;
  M. Bouhmadi-Lopez and  P. Moniz, Phys.Rev. D71 (2005) 063521;
  J.C. Fabris, S.V.B. Goncalves and M.S. dos Santos,
  Gen.Rel.Grav. 36 (2004) 2559;
  M. Biesiada, W. Godlowski and M. Szydlowski, Astrophys.J. 622
  (2005) 28;
  T.Multamaki, M. Manera and E.Gaztanaga, Phys.Rev. D69 (2004)
  023004;
  J. Fabris, S. Goncalves and R. Ribeiro, Gen.Rel.Grav. 36
  (2004) 211;
  M. Szydlowski and W. Czaja, Phys.Rev. D69 (2004) 023506;
  J.  Cunha, J.  Alcaniz and J.  Lima, Phys.Rev. D69 (2004) 083501.
\bibitem{gcg}
  M.C. Bento, O. Bertolami and A.A. Sen, Phys. Rev. {\bf D66}
 (2002) 043507;
\bibitem{sn}
 M. Makler, S.Q. de Oliveira and I. Waga, Phys. Lett.
 {\bf B555} (2003) 1;
 J.C. Fabris, S.V.B. Goncalves and P.E. de Souza,
 astro-ph/0207430;
 Y. Gong and C.K. Duan, Class. Quant. Grav. {\bf 21}
 (2004) 3655; Mon. Not. Roy. Astron. Soc. {\bf 352}
 (2004) 847;
 Y. Gong, JCAP {\bf 0503} (2005) 007.
\bibitem{cmb}
 M.C. Bento, O. Bertolami and A.A. Sen, Phys. Rev. {\bf D67}
 (2003) 063003.
 \bibitem{antiudm}
  H. Sandvik, M. Tegmark, M. Zaldarriaga, I. Waga, Phys.Rev.
  {\bf D69} (2004) 123524, astro-ph/0212114.
 \bibitem{bbc}
  M.~C.~Bento, O.~Bertolami and A.~A.~Sen,
  Phys.\ Rev.\ D {\bf 70}, 083519 (2004)
  ,astro-ph/0407239;
  P.~P.~Avelino, L.~M.~G.~Beca, J.~P.~M.~de Carvalho and C.~J.~A.~Martins,
  JCAP {\bf 0309}, 002 (2003),astro-ph/0307427;
   N.~Bilic, R.~J.~Lindebaum, G.~B.~Tupper and R.~D.~Viollier,
  JCAP {\bf 0411}, 008 (2004).

 \bibitem{zhzh}
  Z. Zhu, Astron.Astrophys. 423 (2004) 421, astro-ph/0411039.
 \bibitem{will}
  C.~M.~Will,
  Living Rev.\ Rel.\  {\bf 4}, 4 (2001)
  ,gr-qc/0103036.


 \bibitem{quph}
  W. Zimdahl, D. Pav¨®n and L. P. Chimento, Phys.Lett. B521 (2001)
 133, astro-ph/0105479;
  Z.K. Guo, R.G. Cai and Y.Z. Zhang, JCAP {\bf 0505} (2005) 002,
  astro-ph/0412624;
  Z.K. Guo and Y.Z. Zhang, Phys.Rev. {\bf D71} (2005) 023501,
  astro-ph/0411524.
 \bibitem{vari}
  U.~Alam, V.~Sahni, T.~D.~Saini and A.~A.~Starobinsky,
  Mon.\ Not.\ Roy.\ Astron.\ Soc.\  {\bf 354}, 275 (2004)
  astro-ph/0311364;
  U.~Alam, V.~Sahni and A.~A.~Starobinsky,
  JCAP {\bf 0406}, (2004) 008,
  astro-ph/0403687;
  D. Huterer and A. Cooray, astro-ph/0404062;
  Y. Wang and M. Tegmark, astro-ph/0501351.




\bibitem{zhzh2}
  Z. Zhu,M. Fujimoto and X. He, Astron.Astrophys. 417 (2004) 833,
astro-ph/0401095.
\bibitem{call}
  R.R. Caldwell, Phys.Lett. B545 (2002) 23, astro-ph/9908168;
  P.~Singh, M.~Sami and N.~Dadhich,
  Phys. Rev. {\bf D68} (2003) 023522, hep-th/0305110.
\bibitem{phantom}
  S. M. Carroll, A. De Felice and M. Trodden,
  Phys.Rev. D71 (2005) 023525, astro-ph/0408081;
  R.~G.~Cai and A.~Wang, JCAP {\bf 0503} (2005) 002,
  hep-th/0411025;
  Z.K. Guo, Y.S. Piao and Y.Z. Zhang,
  Phys.Lett. B594 (2004) 247, astro-ph/0404225;
  S. Nesseris and L. Perivolaropoulos, Phys.Rev. D70 (2004) 123529,
astro-ph/0410309;
  H. Wei and R.G. Cai, hep-th/0501160; H. Wei and R. Cai,
  astro-ph/0509328;
  Y.H. Wei, gr-qc/0502077;
  S.~Nojiri, S.~D.~Odintsov and S.~Tsujikawa,
  Phys.\ Rev.\ D {\bf 71}, 063004 (2005), hep-th/0501025;
  P.~Singh, gr-qc/0502086.
  H.~Stefancic, astro-ph/0504518;
  and referenecs therein.
\bibitem{Vik}A.~Vikman,
  Phys.\ Rev.\ D {\bf 71}, 023515 (2005)
  astro-ph/0407107.
\bibitem{guozk2}
 B.~Feng, X.~L.~Wang and X.~M.~Zhang,
 Phys.\ Lett.\ B {\bf 607}, 35 (2005)
 astro-ph/0404224;
 Z.K. Guo, Y.S. Piao, X. Zhang and Y.Z. Zhang,
 Phys.Lett. B608 (2005) 177, astro-ph/0410654.

\bibitem{And}
  A.~A.~Andrianov, F.~Cannata and A.~Y.~Kamenshchik,
  Phys.\ Rev.\ D {\bf 72}, 043531 (2005)
  ,gr-qc/0505087;
   I.~Y.~Aref'eva, A.~S.~Koshelev and S.~Y.~Vernov,
  Phys.\ Rev.\ D {\bf 72}, 064017 (2005)
  ,astro-ph/0507067;
  S.~Nojiri and S.~D.~Odintsov,
  Phys.\ Rev.\ D {\bf 72}, 023003 (2005)
  ,hep-th/0505215;
  B.~McInnes,
  Nucl.\ Phys.\ B {\bf 718}, 55 (2005)
  ,hep-th/0502209;
  R. Cai, H. Zhang and A. Wang, hep-th/0505186.
\bibitem{bondi}
 H. Bondi, Cosmology, Combrige university press, Combrige, UK,
  1960.
  \bibitem{inter}
  N. Bartolo and M. Pietroni, Phys. Rev. D61, 023518 (2000);
  T. Damour, G. W. Gibbons and C. Gundlach , Phys. Rev. Lett. 64, 123 (1990).

  \bibitem{maia}
   J.M.F. Maia and J.A.S. Lima,
Phys. Rev. D60, 101301(1999).
  \bibitem{stringcos}
 for reviews, see,
  T.~Battefeld and S.~Watson,
  hep-th/0510022;
  A.~Pokotilov,
  UMI-31-59245.
 \bibitem{st}
  R.~Curbelo, T.~Gonzalez and I.~Quiros,
  astro-ph/0502141;
  N. Kaloper and K. A. Olive,
  Phys.Rev.D 57, 811 (1998).


  \bibitem{severalg}
 G. Dvali, G. Gabadadze and  M. Porrati, Phys.Lett. B485 (2000) 208, hep-th/0005016;
  G. Dvali and G. Gabadadze, Phys. Rev. D {\bf 63}, 065007 (2001);
  S. Carroll, V. Duvvuri, M. Trodden and M. Turner, Phys.Rev. D70 (2004)
  043528; C. Lanczos, Z. Phys., 73 (1932) 147;
 D. Lovelock,
  J. Math. Phys., 12 (1971) 498.

  \bibitem{card}
  K. Freese and M. Lewis, Phys.Lett. B540 (2002) 1;
  Z. Zhu and M. Fujimoto, Astrophys.J. 602 (2004) 12,~Astrophys.J. 585, 52(2003)
  ,~Astrophys.J. 581 (2002) 1.






\end{thebibliography}
\end{document}